\documentclass{amsart}
\usepackage{amssymb}
\usepackage{amscd}

 \newtheorem{theorem}{Theorem}
\newtheorem{lemma}[theorem]{Lemma}

\newtheorem{corollary}[theorem]{Corollary}
\newtheorem{definition}{Definition}
\newtheorem{remark}{\it Remark\/}
\newtheorem{example}{\it Example\/}

\DeclareMathOperator{\pr}{pr}
\DeclareMathOperator{\Span}{span}
\DeclareMathOperator{\diag}{diag}
\DeclareMathOperator{\Ad}{Ad}
\DeclareMathOperator{\rank}{rank}
\DeclareMathOperator{\ann}{ann}
\DeclareMathOperator{\ad}{ad}
\newcommand{\R}{\mathbb{ R}}
\newcommand{\h}{\mathfrak{ k}}
\newcommand{\g}{\mathfrak{ g}}
\newcommand{\D}{\mathfrak{ d}}
\newcommand{\A}{\mathrm{ A}}
\newcommand{\T}{\mathbb{ T}}

\newcommand{\F}{\mathcal F}
\newcommand{\I}{\mathrm I\,}
\newcommand{\B}{\mathrm B}

\title[Partial Reductions of Hamiltonian Flows]{Partial Reductions of Hamiltonian Flows and
Hess-Appel'rot Systems on $SO(n)$}

\author{Bo\v zidar Jovanovi\' c}

\begin{document}
\maketitle

\centerline{\small \sc Mathematical Institute SANU}
\centerline{\small \sc Kneza Mihaila 35, 11000 Belgrade, Serbia}
\centerline{\small {\it e-mail}~: bozaj@mi.sanu.ac.yu}

\begin{abstract}
{We study reductions of the Hamiltonian flows restricted to their
invariant submanifolds. As examples, we consider partial
Lagrange-Routh reductions of the natural mechanical systems such
as geodesic flows on compact Lie groups and $n$-dimensional
variants of the classical Hess-Appel'rot case of a heavy rigid
body motion about a fixed point.
\\{\sc MSC:}\, 37J35, 37J15, 53D20}
\end{abstract}

\tableofcontents
\section{Introduction}

In this paper we study reductions of the
Hamiltonian flows restricted to their invariant submanifolds.
Apparently, the lowering of
order in Hamiltonian systems having
invariant relations was firstly studied by Levi-Civita (e.g., see
\cite{LC}, ch. X).

\subsection{Hess-Appel'rot System}
The classical example of the system having an invariant relation
is a celebrated Hess-Appel'rot case of a heavy rigid body motion \cite{He, App}.
Recall that the motion of a heavy rigid body around a fixed point,
in the moving frame, is represented by the Euler-Poisson equations
\begin{equation}
\frac{d}{dt}\vec m=\vec m\times \vec \omega + \mathfrak G \mathfrak M \, \vec \gamma\times \vec r,
\quad \frac{d}{dt}\vec \gamma=\vec \gamma\times \vec \omega, \quad \vec\omega=\A\vec m,
\label{EP}
\end{equation}
where $\vec\omega$ is the angular velocity, $\vec m$ the angular
momentum, $\I=\A^{-1}$ the inertial tensor, $\mathfrak M$ mass and
$\vec r$ the vector of the mass center of a rigid body; $\vec
\gamma$ is the direction of the homogeneous gravitational field
and $\mathfrak G$ is the gravitational constant.

The equations (\ref{EP}) always have three integrals,
the energy, geometric integral and the projection of angular momentum:
\begin{equation}
\F_1=\frac12(\vec m,\vec \omega)+\mathfrak M\mathfrak G(\vec r,\vec
\gamma), \quad  \F_2=(\vec\gamma,\vec\gamma)=1, \quad \F_3=(\vec m,\vec \gamma). \label{integrals}
\end{equation}
For the integrability we need a forth integral. There are three
famous integrable cases: Euler, Lagrange and Kowalevskaya \cite{AKN, Go}.

Apart of these cases, there are various particular solutions
(e.g., see \cite{Go}). The celebrated is partially integrable
Hess-Appel'rot case \cite{He, App}.
Under the conditions:
\begin{equation}
r_2=0, \quad r_1\sqrt{a_3-a_2}\pm r_3\sqrt{a_2-a_1}=0, \quad \A=\diag(a_1,a_2,a_3),
\label{conditions}
\end{equation}
$a_3>a_2>a_1>0$, system (\ref{EP}) has an invariant relation given
by
\begin{equation}
\F_4=(\vec m,\vec r)=r_1 m_1+r_3 m_3=0.
\label{invariant_relation}
\end{equation}

The system is integrable up to one quadrature: the compact connected components of the regular
invariant sets $\F_1=c_1, \F_2=1, \F_3=c_3, \F_4=0$ are tori, but not
with quasi-periodic dynamics. The classical and algebro-geometric integration
can be found in \cite{Go} and \cite{DrGa}, respectively.

There is a nice geometrical
interpretation of the conditions (\ref{conditions}):
the intersection of the plane orthogonal to $\vec r$
with the ellipsoid $(\A\vec m,\vec m)=const$ is a circle
(e.g., see \cite{BoMa}).
If instead of the moving base given by the main axes
of the inertia, we take the moving base $\vec f_1, \vec f_2, \vec
f_3$ such that the mass center of a rigid body $\vec r$ is
proportional to $\vec f_3$,
\begin{equation}
\vec r=\rho \vec f_3, \quad \rho=\sqrt{r_1^2+r_3^2},
\label{r}
\end{equation}
then the inverse of the inertial operator reads
\begin{equation}
\A=\left( \begin{array}{ccc}
a_2 & 0 & a_{13} \\
0 & a_2 & a_{23} \\
a_{13} & a_{23} & a_{33}
\end{array} \right)
\label{HA_operator}
\end{equation}
and the invariant relation (\ref{invariant_relation}) is simply given by
\begin{equation}
m_3=0.
\label{HA_m3}
\end{equation}

Historical overview, with an application  of Levi-Civita ideas to
the Hess-Appel'rot system and other classical rigid body problems
can be found in Borisov and Mamaev \cite{BoMa}.

\subsection{Reduction of Symmetries}
Let $G$ be a Lie group with a free proper Hamiltonian action on a
symplectic manifold $(M,\omega)$. Let
\begin{equation}
\Phi: M \to \mathfrak g^*
\label{moment_map}
\end{equation}
be the corresponding equivariant momentum map. Assume that $\eta$
is a regular value of $\Phi$, so that
\begin{equation}
M_\eta=\Phi^{-1}(\eta)
\label{M_eta}
\end{equation}
and $M_{\mathcal O_\eta}= \Phi^{-1}(\mathcal O_\eta)$ are
smooth manifolds. Here $\mathcal O_\eta=G/G_\eta$ is the coadjoint
orbit of $\eta$. The manifolds $M_\eta$ and $M_{\mathcal O_\eta}$
are $G_\eta$-invariant and $G$-invariant, respectively.

There is a unique symplectic structure
$\omega_\eta$ on $N_\eta=M_\eta/G_\eta$ satisfying
\begin{equation}
\omega\vert_{M_\eta}=d\pi_\eta^*\omega_\eta
\label{symplectic_form}\end{equation}
where $\pi_\eta: M_\eta\to N_\eta$ is the natural projection \cite{MaWe}.

According to Noether's theorem, if $h$ is a $G$-invariant function then the
momentum mapping $\Phi$ is an integral of the Hamiltonian system
\begin{equation}
\dot x=X_h.
\label{Hamiltonian_eq}
\end{equation}
In addition, its restriction to the invariant submanifold  $M_\eta$ projects to the Hamiltonian
system $\dot y=X_H$ on the reduced space $N_\eta$ with $H$
defined by \cite{MaWe}
\begin{equation}
h\vert_{M_\eta}=H\circ \pi_\eta \, .
\label{induced_ham}
\end{equation}

An alternative description of the reduced space is as follows. Let
$\{\cdot,\cdot\}$ be the canonical Poisson bracket on
$(M,\omega)$. Then the manifold $M/G$ carries the induced Poisson
structure and $M_{\mathcal O_\eta}/G$ is the symplectic leaf in
$M/G$. The mapping 
\begin{equation}
\Psi: N_\eta \to M_{\mathcal O_\eta}/G
\label{ISO}
\end{equation} 
which assigns to the $G_\eta$-orbit of $x\in M_\eta$ the $G$-orbit
through $x$ in $M_{\mathcal O_\eta}$ establish the
symplectomorphism between $N_\eta$ and $M_{\mathcal O_\eta}/G$
(e.g., see \cite{OR}).

\subsection{Outline and Results of the Paper}
The aim of this paper is to study Hamiltonian systems that
naturally generalize geometrical properties of the Hess-Appel'rot
system (Definition \ref{def3}, Section 4). This is the reason we
are interested in the following: suppose that $h$ is not a
$G$-invariant function, but (\ref{M_eta}) is still an invariant
manifold of the Hamiltonian system (\ref{Hamiltonian_eq}). As a
modification of the regular Marsden-Weinstein reduction, we study
(partial) reduction of the Hamiltonian system
(\ref{Hamiltonian_eq}) from $M_\eta$ to $N_\eta$ and relationship
between the integrability of the reduced and nonreduced system
(Theorems \ref{main}, \ref{cor}, Section 2).

In Section 3 we define partial Lagrange-Routh reductions (Theorem \ref{LR}). The
construction of the geodesic flows on compact Lie groups,
having invariant manifolds  foliated by invariant tori is presented.
However, the flows over invariant tori are not quasi-periodic.

In Section 4 we consider an $n$-dimensional heavy rigid body
motion about a fixed point. Recall that there are two natural
multidimensional variants of a heavy rigid body which are
described by the Euler-Poisson equations on the dual spaces of
semi-direct products $so(n)\times so(n)$ and $so(n)\times\R^n$.
The integrable generalizations of the Lagrange top with respect to
the first and the second $n$-dimensional variant are given in
\cite{Ra2, DrGa1} and \cite{Be}, respectively. The $n$-dimensional
Kowalevskaya  top in the presence of $q$ gravitational fields
($q\le n$) represented by Euler-Poisson equations on the dual
space of the semi-direct product $so(n)\times (\R^n)^q$ is given
in \cite{RS}.

Recently, an interesting construction of the Hess-Appel'rot system
on $so(n)\times so(n)$ is studied  by Dragovi\' c and Gaji\' c
\cite{DrGa2}. Besides, by generalizing analytical and algebraic
properties of the classical system (\ref{EP}), (\ref{conditions}),
they introduced a class of systems with invariant relations, so
called {\it Hess-Appel'rot type systems} with remarkable property:
there exist a pair of compatible Poisson structures, such that the
system is Hamiltonian with respect to the first structure and
invariant relations are Casimir functions with respect to the
second structure (for more details see \cite{DrGa2}).

We consider the another generalization of a heavy rigid body and
define the $n$-dimensional Hess-Appel'rot system within the
framework of partial reductions (Lemma \ref{l-hess}). It is known
that the motion of a rigid body mass center in the classical
Hess-Appel'rot system is described by the spherical pendulum
equations \cite{Zh, BoMa}. We shall prove the same statement for
the $n$-dimensional variant of the system (Theorem
\ref{pendulum}). On the other side, the L-A pair of the system
written in the closed form of the Euler-Poisson equations on the
dual space of the semi-direct product $so(n)\times\R^n$ is given.
In particular, we get the L-A pair representation for the
Euler-Poisson equations of the classical Hess-Appel'rot system
(\ref{EP}), (\ref{r}), (\ref{HA_operator}) different from those
given in \cite{DrGa} (Theorem \ref{LA-HA}).

Finally, we have made a relationship between our construction and
the Dragovi\' c-Gaji\' c $so(4)\times so(4)$ system (Theorem
\ref{DG}, Section 5).

\section{Partial Reductions and Integrability}

Let $G$ be a compact connected Lie group with a free Hamiltonian action on a
symplectic manifold $(M,\omega)$ with the momentum map (\ref{moment_map}).
Assume that $\eta$ is a regular value of $\Phi$.

\begin{theorem} 
\label{main} {\rm (i)} Suppose that the restriction of $h$ to
$M_{\mathcal O_\eta}$ is a $G$-invariant function.
Then $M_\eta$ is an invariant manifold of
the Hamiltonian system (\ref{Hamiltonian_eq}) 
and $X_{h}\vert_{M_\eta}$ projects
to the Hamiltonian vector field $X_H$
\begin{equation}
d\pi_\eta(X_h)\vert_x=X_H\vert_{y=\pi_\eta(x)},
\label{projection1}
\end{equation}
where $H$ is the induced function on $N_\eta$ defined by (\ref{induced_ham}).

{\rm (ii)}  The inverse statement also holds: if (\ref{M_eta}) is an
invariant submanifold of the Hamiltonian system (\ref{Hamiltonian_eq})
then the restriction of $h$ to $M_\eta$ is a $G_\eta$-invariant function and 
$X_{h}\vert_{M_\eta}$ projects  to the Hamiltonian vector field $X_H$ on $N_\eta$,
where $H$ is defined by (\ref{induced_ham}).
\end{theorem}

In both cases, the Hamiltonian vector field $X_h$ is not assumed to be 
$G$-invariant on $M$.  Moreover $X_h\vert_{M_\eta}$ may not be $G_\eta$-invariant as well. It is invariant modulo
the kernel of $d\pi_\eta$, which is sufficient the tools
of symplectic reduction are still applicable.

\begin{definition}\label{def1}{\rm
We shall refer to the passing from $\dot
x=X_h\vert_{M_\eta}$ to
\begin{equation}
\dot y=X_{H} \label{REDUCED}
\end{equation}
as a {\it partial reduction}.
}\end{definition}

If $h$ is $G$-invariant, the
complete integrability of the reduced and original system are
closely related (see \cite{Jo, Zu}). In our case, we have the
following corollary.

\begin{theorem}
\label{cor} {\rm (i)} Suppose that the partially reduced system
(\ref{REDUCED}) is completely integrable, with a complete set of
commuting integrals $F_1,\dots,F_m$, $m=\frac12\dim N_\eta$. Let
$f_1=F_1 \circ \pi_\eta,\dots, f_n=F_n \circ \pi_\eta$. Then
$M_\eta$ is almost everywhere foliated by ($m+\dim
G_\eta$)-dime\-nsional invariant isotropic manifolds
\begin{equation}
\mathcal M_c=\{f_1=c_1,\dots, f_m=c_m \} \label{M}
\end{equation}
of the system (\ref{Hamiltonian_eq}).

{\rm (ii)} In addition, if $\eta$ is a
regular element of $\mathfrak g^*$ and $f_1,\dots,f_m$ can be
extended to the commuting $G$-invariant functions in some
neighborhood of $M_\eta$, then the compact connected components of
invariant manifolds (\ref{M}) are tori. However, in general, the
flow over the tori is not quasi-periodic.
\end{theorem}

\begin{remark}\label{remark}
{\rm If the partially reduced system (\ref{REDUCED}) is integrable
then we need ($\dim G_\eta$)-additional quadratures for the
solving of $\dot x=X_h\vert_{M_\eta}$ ({\it the reconstruction
equations}). Also, if $G_\eta=G$, i.e., $\mathcal
O_\eta=\{\eta\}$, then $m+\dim G_\eta=\frac12\dim M$. Whence, the
invariant manifolds (\ref{M}) are Lagrangian. }\end{remark}

The partial reduction can be seen as a special case of the
symplectic reductions studied in \cite{BT, Li} (see also
\cite{LM}, Ch. III). There, the {\it symplectic reduction} of a
symplectic manifold $(M,\omega)$ is any surjective submersion $p:
N\to P$ of a submanifold $N\subset M$ onto another symplectic
manifold $(P,\Omega)$, which satisfies $p^*\Omega=\omega\vert_N$.
Also, instead of a compact group action one can consider a proper
group action. However we work in the framework of a regular
Marsden-Weinstein reduction and a compact group action, which
allows us to easily describe the partial reductions of natural
mechanical systems considered in this paper.

\medskip

\noindent{\it Proof of Theorem \ref{main}.} (i) \, Let $\xi_1,\dots,\xi_n$ be the base of $\g$, $(\eta,\xi_\alpha)=\eta_\alpha$ and
\begin{equation}
\phi_\alpha=(\Phi,\xi_\alpha): M\to\R, \quad \alpha=1,\dots,n.
\label{moments}
\end{equation}
Then the level set of the momentum mapping (\ref{M_eta}) is given by the equations
$$
\phi_\alpha=\eta_\alpha, \quad \alpha=1,\dots,n.
$$
The action of $G$ is generated by Hamiltonian vector fields
$X_{\phi_\alpha}$. Since $h\vert_{M_{\mathcal O_\eta}}$ is
$G$-invariant, for $x\in M_\eta$ we have
\begin{equation}
(dh,X_{\phi_\alpha})=\{h,\phi_\alpha\}=-(d\phi_\alpha,X_h)=0,
\quad \alpha=1,\dots,n. \label{INVARIANT-EQ}
\end{equation}
Thus $M_\eta$ is an invariant submanifold.

Let $h^*$ be an arbitrary $G$-invariant function that coincides
with $h$ on $M_{\mathcal O_\eta}$. Then $X_{h^*}$ is a $G_\eta$-invariant
vector field on $M_\eta$ which project to $X_H$ \cite{MaWe}:
\begin{equation}
d\pi_\eta(X_{h^*})\vert_x=X_H\vert_{y=\pi_\eta(x)}.
\label{projection2}
\end{equation}

Let $\delta=h-h^*$. From the condition $\delta\vert_{M_\eta}=0$
we can express $\delta$, in a an open neighborhood of $M_\eta$, as
$$
\delta(x)=\sum_{\alpha=1}^n\delta_\alpha(x)\left(\phi_\alpha(x)-\eta_\alpha\right).
$$

Now, let $f$ be a $G$-invariant function on $M$. Since
$\{f,\phi_\alpha\}=0$ (Noether's theorem), we get
$$
\{\delta,f\}\vert_{M_\eta}= \left(\sum_\alpha
\{\delta_\alpha,f\}\left(\phi_\alpha-\eta_\alpha\right)+
\sum_\alpha \delta_\alpha\{\phi_\alpha,f\}\right)\vert_{M_\eta}=0.
$$
Thus the Poisson bracket $\{\delta,f\}\vert_{M_\eta}=-(df,X_\delta)\vert_{M_\eta}$ vanish for
an arbitrary $G$-invariant function $f$. In other words
\begin{equation}
X_\delta\vert_x \in T_x (G\cdot x)\cap T_x M_\eta \, .\label{*}
\end{equation}

Combining (\ref{projection2}), $X_h=X_{h^*}+X_\delta$ and (\ref{*}) with the well known identity (e.g, see
\cite{OR})
$$
T_x (G\cdot x)\cap T_x M_\eta = T_x (G_\eta \cdot x)=\ker
d\pi_\eta\vert_x\, ,
$$
we prove the relation (\ref{projection1}).

(ii)\,  If $G$ is a connected compact group, the coadjoint isotropy
group $G_\eta$ is connected. Thus, the function $h$ is
$G_\eta$-invariant if and only if it is invariant with respect to
the infinitesimal action of $G_\eta$.

Suppose that $M_\eta$ is an invariant submanifold of
(\ref{Hamiltonian_eq}). Then (\ref{INVARIANT-EQ}) holds for $x\in M_\eta$.
Since the action of $G$ is generated by Hamiltonian
vector fields $X_{\phi_\alpha}$, from (\ref{INVARIANT-EQ}) we get
that $h$ is invariant with repsect to the infinitesimal action of $G_\eta$.
Therefore we have well defined reduced Hamiltonian function $H$ on $N_\eta$.

By using the diffeomorphism (\ref{ISO}) and the fact that $M_{\mathcal O_\eta}$ is a closed
submanifold of $M$, we can find a $G$-invariant function $h^*$ on
$M$ which coincides with $h$ on $M_\eta$. Now, the relation
(\ref{projection1}) follows from the proof of item (i). $\Box$

\medskip

\noindent{\it Proof of Theorem \ref{cor}.} (i) \, Consider a regular
invariant Lagrangian submanifold
\begin{equation}
 \mathcal N_c=\{F_1=c_1,\dots,F_m=c_m \}\subset N_\eta,
\label{NN}
\end{equation}
of the partially reduced system (\ref{REDUCED}). From the
relations (\ref{symplectic_form}) and  (\ref{projection1}) we get
that $\mathcal M_c=\pi_{\eta}^{-1}(\mathcal N_c)$ is an invariant
isotropic manifold of the system (\ref{Hamiltonian_eq}).

(ii) \, Let $\eta\in\g^*$ be a regular element ($G_\eta \approx \T^r$ is a maximal
torus of $G$, $r=\rank\, G$). If the connected component $\mathcal N^o_c$ of
(\ref{NN}) is compact then, by Liouville's theorem, it is
diffeomorphic to a $m$-dimensional torus $\T^m$ with quasi-periodic flow
of (\ref{REDUCED}). Thus the compact connected component
$\mathcal M^o_c=\pi^{-1}_\eta(\mathcal N^o_c)$ of (\ref{M}) is a {\it
torus bundle} over $\T^m$:
\begin{equation}
\begin{array}{cccc}
\T^r & \longrightarrow  & \mathcal M^o_c &  \\
&  & \downarrow  & \pi_\eta  \\
&  &  \T^m &
\end{array}
\label{BUNDLE}\end{equation}

Let $I_1,\dots,I_r$, be the basic
$\Ad^*_G$-invariant polynomials on $\g^*$ and $P_1,\dots,P_{n-r}$
be linear functions on $\g^*$ such that $P_k$, $I_\alpha$ are
independent at $\eta$. Also, let $i_\alpha=I_\alpha \circ \Phi$
and $p_k=P_k\circ \Phi$  be the pull-backs of $I_\alpha$ and $P_k$
by the momentum mapping,  $\alpha=1,\dots,r$, $k=1,\dots,n-r$.

Suppose that $f_1,\dots,f_m$ can be extended to commuting
$G$-invariant functions in some $G$-invariant neighborhood $V$ of
$\mathcal M^o_c$. Then, within $V$, $\mathcal M^o_c$ is given by the
equations
$$
f_1=c_1,\dots, f_m=c_m,  \; i_1= I_1(\eta),\dots, i_{r}=I_r(\eta),
\; p_1=P_1(\eta),\dots,P_{n-4}(\eta)\; .
$$

From the Noether theorem the functions $i_\alpha$, $p_k$ commute
with all $G$-invariant functions on $M$ and, since $i_\alpha$ are
$G$-invariant, the following commuting relations hold on $V$:
\begin{eqnarray}
&\{f_a,f_b\}=\{f_a,i_\alpha\}=\{i_\alpha,i_\beta\}=\{f_a,p_k\}=\{i_\alpha,p_k\}=0,\label{commuting}\\
& a,b=1,\dots,m, \quad \alpha,\beta=1,\dots,r,\quad
k=1,\dots,n-r\, .\nonumber
\end{eqnarray}

Now, as in the case of non-commutative integrability of
Hamiltonian systems \cite{FT, N}, $\mathcal M^o_c$
is a torus  with tangent space spanned by $X_{f_a}$, $X_{i_\alpha}$.
Namely, from the vanishing of the Poisson brackets
(\ref{commuting}) we get that the vector fields $X_{f_a}$,
$X_{i_\alpha}$ are tangential to $\mathcal M^o_c$. Since they are
independent, from the dimensional reasons, they span the tangent
spaces $T_x \mathcal M^o_c$, $x\in \mathcal M^o_c$. Taking into
account (\ref{commuting}) and the relations
$\omega(X_f,X_g)=-\{f,g\}$, $[X_f,X_g]=X_{\{f,g\}}$,  $f,g\in
C^\infty(M)$, we (re)obtain that $\mathcal M^o_c$ is isotropic.
Furthermore the vector fields $X_{f_a}$, $X_{i_\alpha}$ commute
between themselves. Since $\mathcal M^o_c$ is a compact manifold
admitting $m+r=\dim\mathcal M^o_c$ independent commuting vector
fields it is a $(m+r)$-dimensional torus, i.e., the bundle (\ref{BUNDLE})
is trivial.

In general, the flow of $\dot x=X_h$ over the torus $\mathcal M^o_c$
is not quasi-periodic: the vector field $X_{h}$ do not commute
with vector fields $X_{f_1}, \dots, X_{f_m},X_{i_1},\dots,X_{i_r}$
(although Poisson brackets $\{h,f_a\}$, $\{h,i_\alpha\}$ vanish on
$\mathcal M^o_c$). $\Box$

\begin{remark}
{\rm Let $x\in\mathcal M^o$ and $z=\pi(x)$, where $\pi: M\to M/G$ is the canonical projection.
By the use of (\ref{ISO})
and the existence of local  canonical coordinates on the Poisson manifold $M/G$
within some neighborhood  $U$ of $z$ (e.g, see \cite{LM}), the functions $f_i$ can be always extended to $G$-invariant
commuting functions in the $G$-invariant neighborhood $V=\pi^{-1}(U)$.
Thus, if $\mathcal M^o$ is "small enough", it is a torus.
}\end{remark}

\begin{remark}{\rm
It is clear that complete commutative integrability of the reduced system in Theorem \ref{cor},
can be replaced by the condition of non-commutative integrabilty.
}\end{remark}

\begin{remark}{\rm
The Hamiltonian flow (\ref{Hamiltonian_eq}) preserves the
canonical measure $\Omega=\omega^{\dim M/2}$ (the Liouville
theorem). If $M_\eta$ is an invariant manifold of the equations
(\ref{Hamiltonian_eq}), then the functions (\ref{moments}) are
particular integrals. Therefore the time derivative of
$\phi_\alpha$ is of the form
\begin{equation*}\dot \phi_\alpha(x)=\{\phi_\alpha,h\} =\sum_{\beta} \psi_{\alpha\beta}(x)(\phi_\alpha-c_\alpha),
\label{dot_g}
\end{equation*}
where $\psi_{\alpha\beta}$ are smooth functions. Let
$\mathrm{tr}\,\psi$ be the trace of the matrix
$\psi_{\alpha\beta}$. After straightforward calculations, one can
prove that the flow (\ref{Hamiltonian_eq}) restricted to the
invariant manifold $M_\eta$ preserve the restriction of $\Omega$
to $M_\eta$ if and only if
$$
\mathrm{tr}\,\psi(x)=0,
$$ for $x\in
M_\eta$. In particular, if the Hamiltonian $h$ is a $G$-invariant
function, then $\Omega\vert_{M_\eta}$  is an invariant volume
form. }\end{remark}

\section{Partial Lagrange-Routh Reductions}

Let $(Q,l)$ be a  natural mechanical system with
Lagrangian $l=\frac12(\kappa_q \dot q,\dot q)-v(q)$, where the
metric $\kappa$ is also regarded as a mapping $\kappa: TQ\to T^*Q$. The motion of the system is described by the
Euler-Lagrange equations
\begin{equation}
\frac{\partial l}{\partial q} - \frac{d}{dt}\frac{\partial
l}{\partial \dot q}=0 \label{Lagrange}
\end{equation}
or by the Hamiltonian equations on the cotangent
bundle $T^*Q$ with the Hamiltonian
$h(q,p)=\frac12(p,\kappa^{-1}p)+v(q)$ being  the Legendre
transformation of $l$.

Let $G$ be a compact connected Lie group acting freely on $Q$ and
$\pi: Q\to B=Q/G$ be the canonical projection. The $G$-action can
be naturally extended to the Hamiltonian action on $T^*Q$: $g\cdot
(q,p)=(g\cdot q, (dg^{-1})^*p)$ with  the momentum mapping $\Phi$
given by (e.g, see \cite{LM})
$$
(\Phi(q,p),\, \xi)=(p,\, \xi_q), \quad\xi\in\g.
$$
Here $\xi_q$ is the vector given by the action of one-parameter
subgroup $\exp(t\xi)$ at $q$.

For Lagrangian systems, it is convenient to work with tangent
bundle reductions.
Let $\mathcal V_q=\{\xi_q \; \vert \;
\xi\in\g\}$ be  the tangent space to the fibber $G \cdot q$ ({\it
vertical space at} $q$) and $\mathcal V=\cup_q \mathcal V_q$ be
the vertical distribution. Then
$$
(T^*Q)_0=\Phi^{-1}(0)=\cup_q \ann \mathcal V_q, \qquad \ann
\mathcal V_q=\{p\in T^*_q Q\, \vert\, (p,\xi_q)=0, \, \xi\in \g\}.
$$

Consider the {\it horizontal distribution} $\mathcal H=\cup_q
\mathcal H_q\subset TQ$ orthogonal to $\mathcal V$ with respect to
the metric $\kappa$. Equivalently, $\mathcal H$ is the zero level-set of
the tangent bundle momentum mapping $\Phi_l$:
\begin{equation}
\mathcal H=\Phi_l^{-1}(0), \quad (\Phi_l(q,\dot q)\, ,\, \xi)=
\left(\frac{\partial{l}}{{\partial \dot q}}\, ,\, \xi_q\right)=(\kappa_q \dot q\, ,\, \xi_q), \quad \xi\in\g.
\label{horizontal} \end{equation}
Since $\kappa_q(\mathcal H_q)=\ann \mathcal
V_q$, we see that $\mathcal H$ is invariant with respect to the
"twisted" $G$-action
\begin{equation}
{g}\diamond(q,\, X)= (g\cdot q,\, \kappa_{g\cdot q}^{-1}\circ
(dg^{-1})^* \circ \kappa_q (X)), \quad X\in T_q Q, \label{action}
\end{equation}
that is the pull-back of canonical symplectic $G$-action on $T^*Q$
via metric $\kappa$:
\begin{equation*}
\begin{array}{ccccc}
& & \kappa & & \\
& TQ & \longrightarrow  & T^*Q & \\
{g}\diamond &  \downarrow  & &  \downarrow & g\cdot  \\
& TQ & \longrightarrow & T^*Q & \\
 & & \kappa & &
\end{array}
\end{equation*}

From Theorem \ref{main} we obtain

\begin{theorem} \label{LR}
{\rm (i)} The horizontal distribution (\ref{horizontal}) is an invariant
submanifold of the Euler-Lagrange equations (\ref{Lagrange}) if
and only if the potential $v$ and the restriction $\kappa_\mathcal
H$ of the metric $\kappa$ to $\mathcal H$ are $G$-invariant with
respect to the action (\ref{action}).

{\rm (ii)} If $\mathcal H$ is an
invariant submanifold of the system $(Q,l)$ then the trajectories
$q(t)$ of the natural mechanical system with velocities $q(t)$
that belong to $\mathcal H$ project to the trajectories
$b(t)=\pi(q(t))$ of the natural mechanical system $(B,L)$ with the
potential $V(\pi(q))=v(q)$ and the metric $K$  obtained from
$\kappa_\mathcal H$ via identification $\mathcal H/G\approx TB$.
\end{theorem}

Note that when $\kappa$ is $G$-invariant, the twisted $G$-action
(\ref{action}) coincides with usual $G$-action: $g\cdot (q,X)=(g\cdot q, dg (X))$
and the induced metric $K$ is the submersion
metric. In this case Theorem \ref{LR} is exactly the classical method of E. J. Routh 
for eliminating cyclic coordinates \cite{Ro, AKN, MRS}.

\begin{definition}{\rm
By the analogy with the Lagrange-Routh reduction we shall call the procedure of
passing from the Lagrangian system $(Q,l)$ to the system $(B,L)$ a
{\it partial Lagrange-Routh reduction}.}
\end{definition}

\begin{remark}{\rm
The distribution (\ref{horizontal}), in general, is nonintegrable. One can
interpret the restriction of the Lagrangian system $(Q,l)$ to
$\mathcal H$ as a nonholonomic system $(Q,l,\mathcal H)$ with a
property that the reaction forces are equal to zero. }\end{remark}

\subsection{Geodesic Flows on Compact Lie Groups}
Let $G$ be a compact Lie group, $\mathfrak g$ be the Lie algebra of $G$
and $\langle\cdot,\cdot\rangle$ be a $\Ad_G$-invariant scalar product on $\mathfrak g$.
Let $a\in\mathfrak g$ be an arbitrary element and let $b$ belongs  to the center of
$\mathfrak g_a$. Here $\mathfrak g_a=\{ \eta\in \g, [a,\eta]=0\}$ is the isotropy algebra of $a$.
Let $\mathfrak g=\g_a+\mathfrak d$ be the orthogonal decomposition.
Consider the linear operator (so called {\it sectional operators} \cite{FT})
$\A_{a,b,\mathrm C}:\mathfrak g\to \mathfrak g$, defined by
$$
\A_{a,b,\mathrm C}(\xi)=\ad^{-1}_{a} \circ \ad_{b} \circ \pr_{\mathfrak d} (\xi) + \mathrm C (\pr_{\mathfrak g_a} \xi),
$$
where $\pr_{\mathfrak d}$ and $\pr_{\g_a}$  are the orthogonal
(with respect to $\langle\cdot,\cdot \rangle$) projections to
$\mathfrak d$ and $\mathfrak g_a$, respectively and $\mathrm C:
\mathfrak g_a\to \mathfrak g_a$ is symmetric. We can always find
$b$ and $\mathrm C$ such that $\A_{a,b,\mathrm C}$ is positive
definite.

Identify $\mathfrak g^*$ with $\mathfrak g$ by means of the scalar
product $\langle\cdot,\cdot\rangle$ and consider the
left-trivialization:
$$
T^*G \approx_l G\times \mathfrak g=\{(g,\xi)\}.$$
Then the quadratic form
\begin{equation}
h_{a,b,\mathrm C}=\frac{1}{2}\langle\A_{a,b,\mathrm C}(\xi),\xi\rangle
\label{hab}
\end{equation}
can be regarded as the Hamiltonian of
a left-invariant Riemannian metric on $G$. Denote this metric by $\kappa_{a,b,\mathrm C}$.

Let $G_a$ be the adjoint isotropy group of the element $a$ and consider the {\it right} $G_a$-action on $G$.
With the above notation, the momentum mapping and its zero level-set are given by
\begin{eqnarray}
&&\Phi(g,\xi)=\pr_\mathfrak {g_a} (\xi), \label{b-map}\\
&& (T^*G)_0 \approx_l G\times \mathfrak d\,. \label{b-zero}
\end{eqnarray}

\begin{lemma}
The metric $\kappa_{a,b,\mathrm C}$ is invariant with respect to
the $G_a$-action, i.e., the momentum map (\ref{b-map}) is
preserved along the geodesic flow
\begin{equation}
\dot \xi=[\xi,\A_{a,b,\mathrm C}(\xi)], \quad \dot g=g\cdot \A_{a,b,\mathrm C}(\xi)
\label{b-flow}
\end{equation}
if and only if the
quadratic form  $\langle \xi,\mathrm C(\xi)\rangle$ is
$\Ad_{G_a}$-invariant:
\begin{equation}
[\xi,\mathrm C (\xi)]=0, \quad \xi\in \g_a.
\label{C}\end{equation}
\end{lemma}

If $a$ is regular element of the Lie algebra $\g$, then $\g_a$ is Abelian and the condition
(\ref{C}) is satisfied for an arbitrary operator $\mathrm C$.

Suppose (\ref{C}) holds. Then we can project the metric $\kappa_{a,b,\mathrm C}$
to the homogeneous space $G/G_a$, that is to the adjoint orbit of $a$:
$G/G_a\approx O(a)=\{\Ad_g(a)\,\vert\,g\in G\}$.
Since we deal with the right action, the vertical distribution is left-invariant: $\mathcal V_g=g\cdot \mathfrak g_a$,
while from the definition of $\kappa_{a,b,\mathrm C}$, the horizontal distribution is $\mathcal H_g=g\cdot\mathfrak d$
and the submersion metric does not depend on $\mathrm C$.

Denote the submersion metric by $K_{a,b}$. The cotangent bundle
$T^*\mathcal O(a)$ can be realized as a submanifold of
$\mathfrak g\times \mathfrak g$
\begin{equation}
T^*\mathcal O(a)=\{(x,p)\, \vert\, x=\Ad_g(a), p\in\mathfrak
g_x^\perp\}, \label{T*O}\end{equation} with the pairing between
$p\in T^*_x \mathcal O(a)$ and $\eta\in T_x\mathcal O(a)$ given by
$p(\eta)=\langle p,\eta\rangle$. Let $b_x=\Ad_g b$. Then the
Hamiltonian and the geodesic flow for the metric $K_{a,b}$ in
redundant variables $(x,p)$ are given by (see \cite{BJ}) \begin{eqnarray}
&&H_{a,b}(x,p)=\frac12\langle \ad_{b_x} p,\ad_x p\rangle=-\frac12\langle \ad_x\ad_{b_x}p,p\rangle, \label{red_ham} \\
&& \dot x=-\ad_x\ad_{b_x} p=[[b_x,p],x], \label{adjoint1} \\
&& \dot p=-\ad_x^{-1}[p,[x,[b_x,p]]]+\pr_{\g_x}[[b_x,p],p]. \label{adjoint2}
\end{eqnarray}

Now, let us perturb the metric $\kappa_{a,b,\mathrm C}$ as
follows. Take $\delta=\langle \B_\delta (\pr_{\mathfrak d}
\xi),\xi \rangle + \frac12 \langle \mathrm C_\delta (\pr_{\g_a}
\xi),\xi\rangle$, $\B_\delta: \mathfrak  d \to \g_a$, $\mathrm
C_\delta: \g_a \to \g_a$, such that
\begin{equation}
h_\delta(g,\xi)=h_{a,b,\mathrm C}+\delta=\frac12\langle \A_\delta \xi,\xi\rangle \label{perturbation} \end{equation}
is positive definite. Then (\ref{perturbation})
will be the Hamiltonian function of the left-invariant metric that we shall denote by $\kappa_\delta$. Since
$h_\delta=h\vert_{(T^*G)_0}$, the geodesic flow of $\kappa_\delta$
\begin{equation}
\dot \xi=[\xi,\A_{\delta}(\xi)], \quad \dot g=g\cdot \A_{\delta}(\xi).
\label{c-flow}
\end{equation}
has the invariant relation (\ref{b-zero}). Hence we can perform the partial reduction.
The metrics $\kappa_\delta$ and $\kappa_{a,b,\mathrm C}$ induce the same
metric $K_{a,b}$ on the orbit $O(a)$, but their horizontal distributions $\mathcal H^\delta$ and
$\mathcal H$ are different for $\B\ne 0$:
$$
\mathcal H_g^\delta=\kappa_\delta^{-1}(g\cdot \mathfrak d)\ne g\cdot \mathfrak d=\mathcal H_g\,.
$$
In the case when $a$ is a singular element of $\g$ we can take $\delta=\frac12 \langle \mathrm C_\delta (\pr_{\g_a} \xi),\xi\rangle$
such that $\delta$ is not $\Ad_{G_a}$-invariant. Then the perturbed metric has the same horizontal distribution
as the non-perturbed one: $\mathcal H_g^\delta=g\cdot \mathfrak d$.

The geodesic flow (\ref{adjoint1}), (\ref{adjoint2}) is completely
integrable in the non-commutative sense. Moreover, the system is
also integrable in the usual commutative sense by means of
analytic functions, polynomial in momenta and which can be
lifted to commuting $G_a$-invariant functions on $T^*G$ (see \cite{BJ3, MP, BJ}).
Thus, according Theorem \ref{cor} we obtain

\begin{corollary}
The equations (\ref{b-flow}) and (\ref{c-flow}) have the same
invariant isotropic foliation of (\ref{b-zero}).
\end{corollary}

\subsection{Local Description}
To clear up the difference between geodesic flows (\ref{b-flow}) and
(\ref{c-flow}), let us write down the problem in local coordinates.

Suppose $a$ is a regular element of $\mathfrak g$.
Then $G_a\approx \T^n$ ia a maximal torus.
Locally, we have
$T^*G\approx T^*O(a)\times T^*\T^n$. We take coordinates $(q,\varphi,p,\phi)$ in $T^*G$ such that
$$
(q,p)=(q_1,\dots,q_m,p_1,\dots,p_m) \quad {\rm and}\quad (\varphi,\phi)=(\varphi_1,\dots,\varphi_n,\phi_1,\dots,\phi_n)
$$
are canonical local coordinates on $T^*O(a)$ and $T^*\T^n$.
Locally, the right $\T^n$-action  is given by translations in $\varphi$ coordinates and $(T^*G)_0=\{(q,\varphi,p,0)\}$.

The Hamiltonians $h_{a,b,\mathrm C}$ and $h_\delta$ are of the forms:
\begin{eqnarray*}
&& h_{a,b,\mathrm C}=\frac12\sum A^{ij} p_ip_j+\sum B^{i\alpha}p_i\phi_\alpha+\frac12 \sum C^{\alpha\beta}\phi_\alpha \phi_\beta, \\
&&h_\delta=\frac12\sum A^{ij} p_ip_j+\sum B^{i\alpha}_\delta p_i\phi_\alpha+\frac12 \sum C^{\alpha\beta}_\delta \phi_\alpha \phi_\beta,
\end{eqnarray*}
where $A^{ij}$, $B^{i\alpha}$, $C^{\alpha\beta}$ do not depend on the variables $\varphi_\alpha$.

The partially reduced system
\begin{equation}
\dot q_i=\frac{\partial H_{a,b}}{\partial p_i}=\sum A^{ij}p_j,\quad
\dot p_i=-\frac{\partial H_{a,b}}{\partial q_i}=-\sum \frac{\partial A^{kj}}{\partial q_i} p_k p_j, \quad
i=1,\dots,m
\label{red_system}
\end{equation}
is completely integrable and we can treat $q(t)$ and $p(t)$ as known functions of time. 
Here  $H_{a,b}=\frac12\sum A^{ij}p_ip_j$ is the reduced Hamiltonian (\ref{red_ham}).

The equations of the geodesic flows (\ref{b-flow}) and (\ref{c-flow}) on (\ref{b-zero})
in variables $q_i,p_i$ are given by (\ref{red_system}). The equations in variables $\varphi_\alpha$, respectively,  are given by:
$$
\dot\varphi_\alpha=\sum B^{i\alpha}(q) p_i \quad {\rm and}
\quad \dot\varphi_\alpha=\sum B^{i\alpha}_\delta (q,\varphi) p_i, \quad \alpha=1,\dots,n.
$$
While the first system is solvable, the second one, generically,
is not. Globally,  (\ref{b-zero}) is foliated on invariant tori
with quasi-periodic and non-quasi-periodic flows of (\ref{b-flow}) and (\ref{c-flow}), respectively.


\section{$n$-Dimensional Hess-Appel'rot System}

Consider the motion of an $n$-dimensional rigid body around a fixed point $O=(0,0,\dots,0)$ in the
$n$-dimensional Euclidean vector space $(\R^n,(\cdot,\cdot))$.
The configuration space of the system is the Lie group $SO(n)$:
the element $g\in SO(n)$ maps the moving coordinate system (attached to the body)
to the fixed one (e.g., see \cite{FeKo}).
Let $F_1,\dots,F_n$, $E_1,\dots,E_n$ and $f_1,\dots,f_n$,
$e_1,\dots,e_n$ be the orthonormal bases attached to the body and
fixed in the space regarded in the space frame and moving frame:
\begin{eqnarray*}
&&E_1=f_1=(1,0,\dots,0,0)^T,\dots,E_n=f_n=(0,0,\dots,0,1)^T, \\
&&E_1=g\cdot e_1,\dots,E_n=g\cdot e_n, \quad F_1=g\cdot f_1,\dots,F_{n}=g\cdot f_n.
\end{eqnarray*}
We can consider the components of the
vectors $e_1,\dots, e_n$ (or $F_1,\dots,F_n$) as redundant
coordinates on $SO(n)$.

For a path $g(t)\in SO(n)$, {\it the angular velocity in the body
frame} and {\it angular velocity in the space frame} are defined
by $\omega(t)=g^{-1}\cdot g(t) \in so(n)$ and $\Omega(t)=\dot
g\cdot g^{-1}= g\cdot \omega\cdot g^{-1}=\Ad_g\omega$,
respectively. From the conditions $0=\dot E_i=\dot g \cdot
e_i+g\cdot \dot e_i$ and $\dot F_i=\dot g\cdot f_i$,
the vectors $e_1,\dots,e_n$ and $F_1,\dots,F_n$ satisfy Poisson
equations
$$
\dot e_i= -\omega \cdot e_i,  \quad
\dot F_i=\Omega\cdot F_i, \quad  i=1,\dots,n.
$$

The kinetic energy of a rigid body is a left-invariant quadratic form
$\frac12\langle {\I}\omega,\omega\rangle$, where ${\I}\,:\, so(n)\to so(n)$
is a non-degenerate {\it inertia operator} and
$\langle X,Y\rangle=-\frac12{\rm tr}(XY)$ denotes the Killing metric on $so(n)$.
For a ``physical'' rigid body, $\I\omega$ has the form
$I\omega+\omega I$, where $I$ is a symmetric $n\times n$ matrix \cite{FeKo}.
We will relax this condition, considering an arbitrary positive definite operator.

Further, suppose that the body is placed in the homogeneous
gravitational force field in the direction $e_n$ and the position
of the center of mass of a rigid body is $\rho f_n$. Then the
potential is $v=\rho \mathfrak G\mathfrak M\,(f_n,e_n)$, where
$\mathfrak G$ is the gravitational constant and $\mathfrak M$ is
the mass of the body. Let $m=\I \omega$ be the {\it
angular momentum in the body frame} and $\A=\I^{-1}$.
By the use of Killing metric we can identify $so(n)$ and
$so(n)^*$. Then the Hamiltonian in the left-trivialization
\begin{equation}
T^*SO(n)\approx_l SO(n)\times so(n)= \{(g,m)\} \label{left}
\end{equation}
reads
$$
h=\frac12\langle m,\A m\rangle+\rho \mathfrak G\mathfrak M\,(f_n,e_n)
$$
and the equations of the system, in
redundant variables $(e_1,\dots,e_n,m)$,  take the form of the
Euler-Poisson equations
\begin{eqnarray} &&\dot m =[m,\omega ]+
\rho \mathfrak G\mathfrak M \,f_n\wedge e_n, \qquad \omega=\A m \label{EPC}\\
&&\dot e_i=-\omega\cdot e_i, \quad i=1,\dots,n.\label{Poisson1}
\end{eqnarray}

\subsection{Invariant Relations}
Consider the orthogonal, symmetric pair decomposition $\h+\D$ of the Lie algebra $so(n)$:
$$
\h=\Span\{f_i \wedge f_j, \, 1\le i<j \le n-1 \}\cong  so(n-1),
\quad \D=\Span \{f_i \wedge f_n, \, 1 \le i \le n-1\}.
$$

Then we can write $h$ as
$$
h=\frac12\langle m_\h,\A_\mathfrak k m_\mathfrak k \rangle +
\langle m_\h, \B m_\D\rangle+\frac12\langle m_\D,\A_\D m_\D \rangle
+\rho \mathfrak G\mathfrak M\,(f_n,e_n),
$$
where
$ \A_\h=\pr_\h \circ \A \circ \pr_\h , \,\A_\D=\pr_\D \circ \A \circ \pr_\D,\, \B=\pr_\h \circ \A \circ \pr_\D,
m_\mathfrak k=\pr_\h m, \, m_\D=\pr_\D m. $

Let $SO(n-1)$ be the subgroup of $SO(n)$ with the Lie algebra $\mathfrak k$. Consider
the {\it right} $SO(n-1)$ action
(rotations of a rigid body "around" the vector $f_n$).
The momentum mapping and its zero-level set are given by
$\Phi=m_\mathfrak k=\pr_{\mathfrak k} m$ and
\begin{equation}
(T^*SO(n))_0=\Phi^{-1}(0)\approx_l \{(g,m)\, \vert \, m_\mathfrak k=0\}=SO(n)\times\D\,. \label{invariant_set}
\end{equation}

From the relations $[\h,\D]\subset \D$, $[\D,\D]\subset\h$, the
orthogonal projections of equations (\ref{EPC}) to $\h$ and $\D$
are given by
\begin{eqnarray}
&&\dot m_\h =[m_\h,\A_\h m_\h]+[m_\h,\B m_\D]+[m_\D,\A_\D m_\D]+[m_\D,\B^{T} m_\h], \label{m_h}\\
&&\dot m_\D =[m_\h,\A_\D m_\D]+[m_\h,\B^T m_\h]+[m_\D,\A_\h m_\h]+[m_\D,\B m_\D]
+\rho \mathfrak G\mathfrak M \,f_n\wedge e_n. \label{m_d}
\end{eqnarray}

Hence, (\ref{invariant_set}) is an invariant submanifold if and only if
$[m_\D,\A_\D m_\D]=0$, i.e.,  $\langle m_\D,\A_\D m_\D\rangle$ is the
$\Ad_{SO(n-1)}$-invariant quadratic form on $\D$.
Since $SO(n)/SO(n-1)$ is a rank one symmetric space,
$\Ad_{SO(n-1)}$-invariant quadratic form on $\D$ is unique (up to multiplication by a constant).
Thus, we come to the following proposition

\begin{lemma}
\label{l-hess} The submanifold (\ref{invariant_set}) is an invariant
set of the Euler-Poisson equations (\ref{EPC}), (\ref{Poisson1}) if and
only if
\begin{equation}
\pr_\D\circ \A\circ\pr_\D=a\mathrm{Id}_\D,
\label{HA_con}
\end{equation}
where $\mathrm{Id}_\D$ is the identity operator and $a\in\R$.
\end{lemma}

Suppose (\ref{HA_con}) holds. In coordinates $m_{ij}=\langle m, f_i \wedge f_j \rangle$,
the invariant set (\ref{invariant_set}) is given by
\begin{equation}
m_{ij}=0, \qquad 1\le i<j\le n-1.
\label{zero}\end{equation}
For $n=3$, after
usual identification $ (\R^3,\times)\cong (so(n),[\cdot,\cdot])$
\begin{equation}
\vec m \longleftrightarrow
\left( \begin{array}{ccc}
0 & -m_3 & m_2 \\
m_3 & 0 & -m_1 \\
-m_2 & m_1 & 0
\end{array} \right),
\label{identification}
\end{equation}
the operator $\A$ take the form (\ref{HA_operator})
and the invariant relation (\ref{zero}) become (\ref{HA_m3}).

Therefore, it is natural to call the rigid body system (\ref{EPC}),
(\ref{Poisson1}), where (\ref{HA_con}) holds,
a {\it $n$-dimensional Hess-Appel'rot system}.

\subsection{Right Trivialization and Reduction to $S^{n-1}$}
In order to describe the partially reduced system we
consider the problem in the right-trivialization
$$T^*SO(n)\approx_r SO(n)\times so(n)=\{(g,M)\},$$
i.e., in the frame fixed in the space. Instead of $(e_1,\dots,e_n)$ we use
$(F_1,\dots,F_n)$ as redundant coordinates on $SO(n)$.
Then the sphere
$S^{n-1}=SO(n)/SO(n-1)$ can be identified with the positions of
the vector $F_n$:
\begin{equation*}
\begin{array}{cccc}
SO(n-1) & \longrightarrow & SO(n)   \\
&  & \downarrow  \pi  \\
&  &  S^{n-1}
\end{array},  \qquad \pi(F_1,\dots,F_n)=F_n\,.
\end{equation*}

The equations of motion (\ref{EPC}), (\ref{Poisson1}) in the space
frame take the form
\begin{eqnarray}
&&\dot M=\rho \mathfrak G\mathfrak M \,F_n\wedge E_n, \label{right_equations}\\ && \dot F_i=\Omega\cdot F_i, \quad i=1,\dots,n, \nonumber
\end{eqnarray}
where the angular momentum $M$ and velocity  $\Omega$ in the space are
related to the angular momentum $m$ and velocity $\omega$ in body coordinates
via $M=\Ad_g m$ and $\Omega=\Ad_g \omega$,
respectively. Therefore
\begin{equation}
\Omega=\Ad_{g} (\omega) =\Ad_{g} \circ \A \circ \Ad_{g^{-1}} (M).
\label{right_operator}
\end{equation}

Since $F_i\wedge F_j=\Ad_g(f_i\wedge f_j)$,
the invariant submanifold $(\ref{invariant_set})$ in the right-trivialization reads
$$
(T^*SO(n))_0 \approx_r\{(g,\mathcal D_g), \, \mathcal D_g= \Ad_g (\D)\}.
$$

Let $\mathcal K_g=\Ad_g(\h)$ and $X=X_{\mathcal K_g}+X_{\mathcal
D_g}$ be the decomposition of $X\in so(n)$ with respect to
$so(n)=\mathcal K_g+\mathcal D_g$. Note that the orthogonal
projection to $\mathcal D_g$  given by
\begin{equation} X_{\mathcal D_g}=(X\cdot F_n)\wedge F_n. \label{projection}
\end{equation}

On $(T^*SO(n))_0$ we have
$\omega=A_\D(m)+\B (m)=a\cdot m+\B(m)$.
Thus
$$
\Omega=\Omega_{\mathcal K_g}+\Omega_{\mathcal D_g}, \quad
\Omega_{\mathcal K_g}=\Ad_g\circ\B\circ\Ad_{g^{-1}}(M), \quad
\Omega_{\mathcal D_g}=a\cdot M,
$$

Furthermore, from (\ref{projection}) we have $\Omega\cdot
F_n=\Omega_{\mathcal D_g} \cdot F_n$. Therefore, on
$(T^*SO(n))_0$, the Hess-Appel'rot system form the closed system
in variables $(F_n,\Omega_{\mathcal D_g})$:
\begin{eqnarray} &&\dot \Omega_{\mathcal D_g}=a{\rho \mathfrak G\mathfrak M}\, F_n\wedge E_n, \label{HA}\\
&& \dot F_n=\Omega_{\mathcal D_g} \cdot F_n, \nonumber
\end{eqnarray}

From the second equation we get
\begin{equation}
\Omega_{\mathcal D_g}=\dot  F_n\wedge F_n, \quad {\rm i.e.,} \quad
M=\frac{1}{a} \dot  F_n\wedge F_n \label{MM}\end{equation} and the
first equation can be rewritten in the form:
\begin{equation}
\ddot F_n=-a{\rho \mathfrak G\mathfrak M}\, E_n+\mu F_n\,.
\label{spherical_pendulum} \end{equation}
Here the Lagrange multiplier $\mu$ is determined from the condition $(F_n,F_n)=1$.

Therefore we obtain:

\begin{theorem}
\label{pendulum} The motion of the center of the mass of the
$n$-dimensional Hess-Appel'rot system
(\ref{EPC}), (\ref{Poisson1}), (\ref{invariant_set}), (\ref{HA_con}) in the space frame is
described by the spherical pendulum equations
(\ref{spherical_pendulum}).
\end{theorem}

The same statement, for the classical Hess-Appel'rot system is
proved by Zhu\-ko\-vski (see \cite{Zh, BoMa}).

It is well known that the partially reduced system (\ref{spherical_pendulum}) 
is completely integrable in the non-commutative sense. The tangent bundle $TS^{n-1}$ is foliated
by two-dimensional invariant manifolds that are level-sets of the
energy and integrals \begin{equation*}
\langle \dot F_n \wedge F_n, E_i\wedge E_j \rangle =(\dot F_n,E_i)(F_n,E_j)-(\dot F_n,E_j)(F_n,E_i), \quad
1\le i<j \le n-1.
\label{int00}
\end{equation*}

Using (\ref{MM}), we see that the lifting of these integrals to
$(T^*SO(n))_0$ are components of the momentum map of the {\it
left} $SO(n-1)$-action:
\begin{equation}
\mathcal G_{ij}=a\cdot  \langle M, E_i\wedge E_j \rangle, \quad 1\le i<j \le n-1
\label{int0}
\end{equation}
(note that (\ref{int0}) are integrals on the whole phase space $T^*SO(n)$ as well). Whence

\begin{corollary} \label{COR}
The invariant set (\ref{invariant_set}) of the Hess-Appel'rot system
is almost everywhere foliated by $(2+(n-1)(n-2)/2)$-dimensional isotropic invariant manifolds, level sets
of the Hamiltonian function and integrals (\ref{int0}).
\end{corollary}

The above considerations motivate us for the following definition

\begin{definition}\label{def3}{\rm
We shall say that a Hamiltonian system (\ref{Hamiltonian_eq})
satisfies {\it geometrical Hess-Appel'rot conditions} if it has an
invariant relation (\ref{M_eta}) and the partially reduced system
(\ref{REDUCED}) is completely integrable. }\end{definition}

The geodesic flows on compact Lie groups presented in Section 3 provide  examples of systems
that satisfy geometrical Hess-Appel'rot conditions.
Note that the condition that the partially reduced system is integrable
differs from the notion of {\it restricted integrability} given in \cite{DrGa2}.

\subsection{Reduction to $(so(n)\times \R^n)^*$}
The system (\ref{EPC}), (\ref{Poisson1}) is always {\it left}
$SO(n-1)$-invariant (rotations of a rigid body "around" the vector
$e_n$). Denote $\gamma=e_n$, $r=\rho f_n$. The equations (\ref{EPC}) together with the last Poisson
equation,
\begin{equation}
\dot m =[m,\omega ]+ \mathfrak G\mathfrak M\, r\wedge \gamma, \quad \dot \gamma=-\omega\cdot \gamma, \quad \omega=\A m, \label{EP2}  \end{equation}
can be seen as a left $SO(n-1)$-reduction of (\ref{EPC}),
(\ref{Poisson1}). This is a Hamiltonian system on the dual space of Lie algebra
$e(n)=so(n)\times\R^{n}$ with respect to the usual Lie-Poisson bracket.
For $n=3$, after identification (\ref{identification}), the equations (\ref{EP2}) get the familiar form (\ref{EP}).

Now suppose the Hess-Appel'rot condition (\ref{HA_con}) is satisfied.
Then the Euler-Poisson equations (\ref{EP2}) on the invariant set (\ref{zero}) coordinately read:
\begin{eqnarray}
&&\dot m_{in}=-\sum_{j=1}^{n-1} \omega_{ij} m_{jn} - \rho\mathfrak G\mathfrak M\, \gamma_i,
\quad i=1,\dots,n-1, \nonumber \\
&&\dot \gamma_i=-a\gamma_n m_{in}-\sum_{j=1}^{n-1} \omega_{ij} \gamma_j, \quad i=1,\dots,n-1, \label{Hess}\\  &&\dot \gamma_n=a \sum_{j=1}^{n-1} \gamma_j m_{jn}, \nonumber
\end{eqnarray}
where $\omega_{ij}=\langle \B m, f_i\wedge f_j\rangle$, $1 \le i<j \le n-1$.
The equations (\ref{Hess}) always have three integrals
\begin{eqnarray}
&&\F_1=\frac{a}2 \sum_{i=1}^{n-1} m_{in}^2 +\rho\mathfrak G\mathfrak M\, \gamma_n,
\quad \F_2=\sum_{i=1}^{n} \gamma_i^2=1, \nonumber \\
&&\F_3=\sqrt{\sum_{1\le i<j \le
n-1}(m_{in}\gamma_j-m_{jn}\gamma_i)^2}, \label{integrals2}
\end{eqnarray}
which correspond to integrals (\ref{integrals}). As we sow above, the
variable $\gamma_n=(e_n,f_n)=(E_n,F_n)$
satisfies the vertical component equation for the motion of the spherical pendulum
and can be found by quadratures.

In general, equations (\ref{Hess}) have no smooth invariant measure.
The non-existence of a smooth invariant measure
is the reflection of the non-solvability by quadratures of the classical system (\ref{EP}),
(\ref{conditions}), (\ref{invariant_relation}).
Namely, by the Euler-Jacobi theorem
(e.g., see \cite{AKN}, page 131),
the invariant measure together with the integrals
(\ref{integrals}), (\ref{invariant_relation}) would
implies solvability of the system by quadratures.

\begin{example}{\rm As an illustration, consider the case $n=4$ and Hamiltonian
\begin{eqnarray*}
h&=&\frac12(a_1 m_{23}^2+a_2 m_{13}^2+a_3 m_{12}^2)+\frac {a}{2}(m_{14}^2+ m_{24}^2 + m_{34}^2)+\\
&& + b_1 m_{12} m_{14} + b_2 m_{12} m_{24} + b_3 m_{12} m_{34}+\rho\mathfrak G\mathfrak M \gamma_n.
\end{eqnarray*}
Then the equations in variables $m_{12}, m_{23}, m_{13}$ are
\begin{eqnarray*}
&& \dot m_{12} =m_{13}m_{23}(a_2-a_1)+ m_{12}(b_1 m_{24}- b_2 m_{14}),    \\
&&  \dot m_{23}= m_{12}m_{13}(a_3-a_2) + m_{12}(b_2 m_{34}- b_3 m_{24})+
m_{13}(b_1 m_{14} + b_2 m_{24} + b_3 m_{34}) ,\\
&&  \dot m_{13}= m_{23}m_{12}(a_1-a_3) + m_{12}(b_1 m_{34}- b_3 m_{14})-
m_{23}(b_1 m_{14} + b_2 m_{24} + b_3 m_{34}),
\end{eqnarray*}
while the equations in other variables, on the invariant set
\begin{equation}
m_{12}=m_{13}=m_{23}=0,
\label{zero2}
\end{equation}
take the form
\begin{eqnarray} &&\dot m_{14}=-m_{24}(b_1 m_{14} + b_2 m_{24} + b_3 m_{34})-\rho\mathfrak G\mathfrak M \gamma_1, \nonumber\\
&&\dot m_{24}=m_{14}(b_1 m_{14} + b_2 m_{24} + b_3 m_{34})-\rho\mathfrak G\mathfrak M \gamma_2, \nonumber\\
&&\dot m_{34}=-\rho\mathfrak G\mathfrak M \gamma_3, \nonumber\\
&&\dot \gamma_1=-a \gamma_4 m_{14}- \gamma_2(b_1 m_{14} + b_2 m_{24} + b_3 m_{34}),\label{Hess2}\\
&&\dot \gamma_2=-a \gamma_4 m_{24} +\gamma_1(b_1 m_{14} + b_2 m_{24} + b_3 m_{34}),\nonumber\\&&\dot \gamma_1=-a \gamma_4 m_{34},\nonumber\\
&& \dot \gamma_4=a(\gamma_1m_{14}+\gamma_2 m_{24} +\gamma_3m_{34}).\nonumber
\end{eqnarray}

Together with (\ref{integrals2}) equations (\ref{Hess2})
have a supplementary integral
\begin{equation}
\F_{12}=m_{14}\gamma_2-m_{24}\gamma_1,
\label{int1}\end{equation}
implying  the foliation on 3-dimensional invariant manifolds of (\ref{zero2}).

In redundant coordinates $(F_1,F_2,F_3,F_4,M)$
the integral (\ref{int1}) on $T^*SO(4)$  reads
\begin{eqnarray}
\F_{12}&=&\langle m,f_1\wedge f_4 \rangle (e_4,f_2)-\langle m,f_2 \wedge f_4 \rangle (e_4,f_1)\label{int1a}\\&=& \langle M,F_1\wedge F_4 \rangle (E_4,F_2)-\langle M,F_2 \wedge F_4 \rangle (E_4,F_1)\,.\nonumber\end{eqnarray}
It is clear that $\F_{12}$ is independent of integrals (\ref{int0}).
Whence the 5-dimensional invariant isotropic manifolds outlined in Corollary \ref{COR}
are foliated on 4-dimensional invariant level-sets of (\ref{int1a}).

Note that the system (\ref{Hess2}) has an invariant measure if and only if $b_1=b_2=0$.
Also, recall that for $a_1=a_2=a_3 \ne a$, $b_1=b_2=b_3=0$, the Euler-Poisson equations represent
the motion of a dynamically symmetric heavy
rigid body, 4-dimensional version of the Lagrange top \cite{Be}.

Therefore, if $b_1=b_2=b_3=0$ and $a_1\ne a_2 \ne a_3$
then the equations (\ref{Hess2}) coincides to the 4-dimensional Lagrange top equations
restricted to the invariant level-set (\ref{zero2}), although the systems are different on the whole phase space.
They have additional integrals
\begin{eqnarray}
&&\mathcal F_{13}=m_{14}\gamma_3-m_{34}\gamma_1 \,
\left(=\langle M,F_1\wedge F_3 \rangle (E_4,F_3)-\langle M,F_3 \wedge F_4 \rangle (E_4,F_1)\right),\nonumber\\
&&\mathcal F_{23}=m_{24}\gamma_3-m_{34}\gamma_2 \,
\left(=\langle M,F_2\wedge F_4 \rangle (E_4,F_3)-\langle M,F_3 \wedge F_4 \rangle (E_4,F_2)\right).
\label{int2}\end{eqnarray}

Since the equations (\ref{Hess2}) preserve the standard measure in variables $m_{14},m_{24}, m_{34}$,
$\gamma_1,\gamma_2,\gamma_3,\gamma_4$ and
the invariant manifolds given by integrals (\ref{integrals2}), (\ref{int1}), (\ref{int2}) are two-dimensional,
by the Euler-Jacobi theorem, they are solvable by quadratures.
On the other side, the Hamiltonian and functions (\ref{int0}), (\ref{int1a}), (\ref{int2}) provide 3-dimensional invariant foliation 
of 9-dimensional manifold $(T^*SO(4))_0$.
}\end{example}

\subsection{L-A Pair}
The non-commutative integrability of the $n$-dimensional Lagrange top is proved by Beljaev \cite{Be} and the L-A pair is given by
Reyman and Semenov-Tian-Shansky \cite{RS}. The L-A pair of the similar form can be used for the Hess-Appel'rot system on $so(n)\times \R^n$.
Let $m^*$, $\omega^*$, $\gamma^*$, $r^*$ be the $so(n+1)$-matrixes given by
\begin{eqnarray*}
&&m^*=
\left( \begin{array}{cc}
m & 0  \\
0 & 0
\end{array} \right), \quad
\gamma^*=
\left( \begin{array}{cc}
\mathbf{0} & \gamma  \\
-\gamma^t & 0
\end{array} \right), \\
&&\omega^*=
\left( \begin{array}{cc}
\omega & 0  \\
0 & 0
\end{array} \right), \quad
r^*=
\left( \begin{array}{cc}
\mathbf{0} & \rho f_n  \\
-\rho f_n^t & 0
\end{array} \right),
\end{eqnarray*}
where $\mathbf{0}$ is the zero $n\times n$ matrix.

Then, under the Hess-Appel'rot conditions (\ref{HA_con}) and (\ref{zero}),
the Euler-Poisson equations (\ref{EP2})
are equivalent to the matrix equation with a spectral parameter $\lambda$
\begin{equation} \dot L(\lambda)=[L(\lambda),A(\lambda)],
\label{LA_pair}
\end{equation}
where
$L(\lambda)=\gamma^*+\lambda  m^*  + \lambda^2 \frac{1}{a} \mathfrak G\mathfrak M r^*$ and
$A(\lambda)=\omega^*+ \lambda \mathfrak G\mathfrak M r^*$.

Indeed, the straightforward computations shows that
the terms with $\lambda^0$ and $\lambda^1$ in (\ref{LA_pair}) are
equivalent to the equations (\ref{EP2}). The  left hand side term
with $\lambda^2$ is identically equal to zero. It can be proved
that the right hand side term with $\lambda^2$ vanish, on the
invariant submanifold (\ref{zero}), if and only if
(\ref{HA_con}) holds.

On the invariant set (\ref{zero}) the spectral curve is
\begin{eqnarray}
&& p(\lambda,\mu)=\det(L(\lambda)-\mu\mathrm{Id})=(-\mu)^{n-3}\left(\mu^4+\mu^2 P(\lambda)+Q(\lambda)^2\right)=0,  \label{CURVE}\\
&& P(\lambda)=\F_2+\frac{2}{a}\lambda^2 \F_1 + \lambda^4
\left(\frac{\rho\mathfrak G\mathfrak M}{a}\right)^2, \quad
Q(\lambda)=\lambda {\mathcal F_3},\nonumber
\end{eqnarray}
where $\F_1,\F_2,\F_3$ are integrals (\ref{integrals2}).
The curve  (\ref{CURVE}) is reducible (for $n>3$)
and consists of the $n-3$ copies of the rational curve $\mu=0$ and the singular algebraic
curve
\begin{equation}
\Gamma: \quad \mu^4+\mu^2 P(\lambda)+Q(\lambda)^2=0.
\label{curve}
\end{equation}

The spectral curves of the form (\ref{curve}) appear in the
algebro-geometric analysis given in \cite{DrGa1, DrGa2}.

In particular the above consideration leads to the L-A pair of the
classical system different from those obtained in \cite{DrGa}.

\begin{theorem} \label{LA-HA}
Under the Hess-Appel'rot conditions (\ref{r}),
(\ref{HA_operator}), (\ref{HA_m3}), the Euler-Poisson equations
(\ref{EP}) are equivalent to the matrix equation (\ref{LA_pair}),
where
\begin{eqnarray*}
&& L(\lambda)=\left( \begin{array}{cccc}
0 & -\lambda m_3 & \lambda m_2 & \gamma_1 \\

\lambda m_3 & 0 & -\lambda m_1 & \gamma_2\\
-\lambda m_2 & \lambda m_1 & 0 & \gamma_3+\lambda^2\mathfrak G\mathfrak M\frac{\rho}{a_2}\\
-\gamma_1  & -\gamma_2 & -\gamma_3-\lambda^2\mathfrak G\mathfrak M\frac{\rho}{a_2} & 0
\end{array} \right), \\&& A(\lambda)=\left( \begin{array}{cccc}
0 & -\omega_3 & \omega_2 & 0  \\
\omega_3 & 0 & -\omega_1 & 0\\
-\omega_2 & \omega_1 & 0 & \lambda \rho \mathfrak G\mathfrak M\\
0  & 0 & -\lambda\rho \mathfrak G\mathfrak M & 0
\end{array} \right).
\end{eqnarray*}
\end{theorem}

\begin{remark}{\rm
Recall that the spectral curve for the classical system given in \cite{DrGa} is reducible
and consists of the rational and the elliptic curve. In our case,
due to the involution $\tau: (\lambda,\mu)\mapsto (\lambda,-\mu)$,
the singular spectral curve (\ref{curve}) is a two-covering of the regular
(for the generic values of the integrals (\ref{integrals})),
genus-3 hyperelliptic curve
$$
\Gamma'=\Gamma/\tau: \, u^2+u P(\lambda)+Q(\lambda)^2=0 $$
(e.g., see Lemma 1 in \cite{DrGa1}). In the affine part, the point $(0,0)\in\Gamma$ is an ordinary
double point as well as the ramification point of the covering.
}\end{remark}

\section{Hess-Appel'rot system on ($so(4)\times so(4))^*$ }

Dragovi\'c and Gaji\'c  defined a class of systems of
the Hess-Appel'rot type by using the analytical and algebraic
properties of the classical system \cite{DrGa2}.
Subtle difference between Dragovi\'c-Gaji\'c systems and those constructed here
has to be studied carefully, but there is a big coherence between these two approaches.
Remarkably, the $so(4)\times so(4)$ system given in \cite{DrGa2}
considered on the whole phase space $T^*SO(4)$ satisfies
geometrical Hess-Appel'rot conditions as well.

We start from the example of the geodesic flows given in Section 3.
Suppose, in addition, the $G_a$-invariant potential force $v(g)$
is given. Then (\ref{b-zero}) remains to be an invariant manifold for
natural mechanical systems with kinetic energies
$\kappa_{a,b,\mathrm C}$ and $\kappa_\delta$. After Lagrange-Routh
and partial Lagrange-Routh reductions, the systems project to the
system with kinetic energy $K_{a,b}$ and potential force $V(x)$ on
the adjoint orbit $\mathcal O(a)$, where $V(x)=v(g)$, $x=\Ad_g a$.

The reduced system is a modification of (\ref{adjoint1}), (\ref{adjoint2}) by the potential force:
\begin{eqnarray}
&& \dot x=-\ad_x\ad_{b_x} p=[[b_x,p],x], \label{adjoint3} \\
&& \dot p=-\ad_x^{-1}[p,[x,[b_x,p]]]-\frac{\partial V(x)}{\partial x} + \sum_{i=1}^n \lambda_i \xi_i(x). \label{adjoint4}
\end{eqnarray}
Here $\xi_1(x),\dots,\xi_n(x)$ is a base of $\g_x$ and Lagrange
multipliers $\lambda_i$ are chosen such that trajectory
$(x(t),p(t))$ belongs to (\ref{T*O}). For $a$ proportional to $b$
and a linear potential $V(x)=\langle x,c\rangle$, the system
represents analog of spherical pendulum on adjoint orbit $O(a)$,
which is integrable on an {\it arbitrary} orbit $O(a)$ (see
Bolsinov and Jovanovi\' c \cite{BJ2}).

Now consider the above construction for the case $G=SO(4)$, i.e, for the $4$-dimensional
rigid body motion about a fixed point (we use notion from the previous section).
Let
\begin{equation}
b=(J_1+J_3) a, \quad a=a_{12} f_1 \wedge f_2 + a_{34} f_3\wedge f_4.
\label{ab}
\end{equation}
We have $so(4)_a=\Span\{ f_1 \wedge f_2, f_3 \wedge f_4\}=so(2)\oplus so(2)$.
Define
$$
\mathrm C: so(4)_a \to so(4)_a, \quad \mathrm C(f_1\wedge f_2)=2J_1 f_1\wedge f_2, \quad
\mathrm C(f_3\wedge f_4)=2J_3 f_3\wedge f_4.
$$

Then the left-invariant kinetic energy $h_{a,b,\mathrm C}$ (see (\ref{hab})) takes the form
$$
h_{a,b,\mathrm C}(g,m)= J_1 m_{12}^2+J_3 m_{34}^2 + \frac{J_1+J_3}{2}(m_{13}^2+m_{14}^2+m_{23}^2+m_{24}^2).
$$
Consider the perturbation of the metric
\begin{eqnarray}
&&h_{\delta}=h_{a,b,\mathrm C}+\delta=\frac12 \langle \A_\delta m,m \rangle,\label{ENERGY} \\
&&\delta(g,m)=m_{12}(J_{14} m_{14}-J_{13}
m_{23})+m_{34}(J_{13}m_{14}-J_{14}m_{23})\nonumber
\end{eqnarray}
and the potential function
\begin{eqnarray}
v &=& \langle \Ad_{g^{-1}}(a), a \rangle=
\langle a_{12} e_1 \wedge e_2 + a_{34} e_3 \wedge e_4, a  \rangle \nonumber \\
&=& \langle a,\Ad_g(a)\rangle= \langle a, a_{12} F_1 \wedge F_2 +
a_{34} F_3 \wedge F_4\rangle \label{POTENTIAL}
\end{eqnarray}
(written in the coordinates $(e_1,e_2,e_3,e_4)$ and
$(F_1,F_2,F_3,F_4)$, respectively).

The rigid body system with Hamiltonian $h=h_\delta+v$, in the left-trivialization (\ref{left}), takes the form
\begin{equation}
\dot m=[m,\A_{\delta}(m)]+[a_{12} e_1\wedge e_2+a_{34} e_3 \wedge e_4,a], \quad \dot g=g\cdot \A_{\delta}(m).
\label{ha1} \end{equation}

The system (\ref{ha1}) has two invariant relations
\begin{equation} m_{12}=0, \quad m_{34}=0 \label{zero_so(4)}\end{equation} and it
is reducible to the {\it oriented Grassmannian variety} $Gr^+(4,2)\approx O(a)$ of oriented
two-dimensional planes in the four-dimensional vector space.
The diffeomorphism $\Psi: O(a) \to Gr^+(4,2)$ is simply given by
$$
\Psi(x)=F_1\wedge F_2, \quad {\rm where} \quad x=\Ad_g(a)=a_{12}
F_1 \wedge F_2 + a_{34} F_3 \wedge F_4.
$$
Recall that $F_1,F_2,F_3,F_4$ is the moving base regarded in the
space frame, so $F_1 \wedge F_2$ is the oriented two-plane
attached to the body.

From (\ref{red_ham}), (\ref{ab}), (\ref{POTENTIAL}) we get the
reduced Hamiltonian
$$
H(x,p)=H_{a,b}(x,p)+V(x)=\frac{J_1+J_3}{2}\langle [x,p],[x,p]\rangle+\langle x,a\rangle\,.
$$
After calculating the Lagrange multipliers in (\ref{adjoint3}), (\ref{adjoint4})
the partially reduced system  on  $T^*Gr^+(4,2)$ become
\begin{eqnarray}
&& \dot x=(J_1+J_3)[[x,p],x], \label{p1}\\
&& \dot p=(J_1+J_3) [[x,p],p] -a + \pr_{\g_x} a.\label{p2}
\end{eqnarray}

It follows from Theorem 4 \cite{BJ2} that the reduced system is
completely integrable by means of integrals that can be extended
to commuting $SO(4)_a$-invariant functions on $T^*SO(4)$. Hence we get
the following qualitative behavior of the system

\begin{theorem} \label{DG}
The rigid body system (\ref{ha1}) satisfies geometrical
Hess-Appel'rot conditions: the partial reduction of the system
from (\ref{zero_so(4)}) to the oriented Grassmannian variety $Gr^+(4,2)$ is
completely integrable pendulum type system (\ref{p1}), (\ref{p2}).
Further, 10-dimensional invariant manifold (\ref{zero_so(4)}) is
almost everywhere foliated by invariant 6-dimensional Lagrangian
tori that project to the 4-dimensional Liouville tori of the
reduced system (\ref{p1}), (\ref{p2}). We need two additional
quadratures to solve the reconstruction problem.
\end{theorem}

By introducing $\gamma=a_{12} e_1 \wedge e_2+a_{34} e_3\wedge e_4$,
from (\ref{ha1})
we can write the closed system in variables $(m,\gamma)$:
\begin{equation}
\dot m=[m,\omega]+[\gamma,a], \quad \dot\gamma=[\gamma,\omega], \quad \omega=\A_{\delta}(m).
\label{ha2}
\end{equation}

Regarding $\gamma$ as the free $so(4)$-variable, the system (\ref{ha2}) becomes the
Hamiltonian system with respect to the Lie-Poisson bracket on the dual space of the semi-direct product
$so(4) \times so(4)$ (see Ratiu \cite{Ra2}). Moreover, with the above choice of $h_\delta$ we have
$$
\A_\delta (m)=Jm+mJ, \quad J=\left(
\begin{array}{cccc}
J_1 & 0 & J_{13} & 0  \\
0 & J_1 & 0 & J_{24} \\
J_{13} & 0 &  J_3 & 0 \\
0 & J_{24} & 0 & J_3
\end{array} \right),
$$
and (\ref{ha2}) coincides with the Hess-Appel'rot
system defined in \cite{DrGa2}.

\begin{remark}{\rm
A similar statement can be proved for the Hess-Appel'rot systems on $so(n)\times so(n)$
for $n>4$ (see \cite{DrGa2}), where one should take
$$
b=(J_1+J_3) a, \quad
a=a_{12} f_1 \wedge f_2.
$$
Then $so(n)_a=so(2)\oplus so(n-2)$ and the systems are
partially reducible to the orbits $O(a)$, which are now
diffeomorphic to the oriented Grassmannian varieties
$Gr^+(n,2)$ of oriented
two-dimensional planes in the $n$-dimensional vector space.}\end{remark}

Relations (\ref{zero_so(4)}) define 6-dimensional invariant manifolds within
generic symplectic leafs of $(so(4)\times so(4))^*$. Appart of the Hamiltonian function,
the system (\ref{ha2}) has a supplementary integral (see \cite{DrGa2})
$$
\F=\gamma_{34}a_{12}+\gamma_{12} a_{34} +(J_1+J_3)(m_{12}m_{34}+m_{23}m_{14}-m_{13}m_{24})
$$
implying the foliation on 4-dimensional invariant varieties. The
L-A pair representation as well as the  algebro-geometric
integration of (\ref{ha2}) up to two quadratures are given in
\cite{DrGa2}.

\subsection*{Acknowledgments} I am very grateful to Vladimir
Dragovi\' c and Borislav Gaji\' c for useful discussions and to the
referees for suggestions which help me to improve the exposition of the paper. The
research was supported by the Serbian Ministry of Science,
Project "Geometry and Topology of Manifolds and Integrable Dynamical Systems".

\end{document}